\documentclass[a4paper]{elsarticle}

\usepackage{graphicx}
\usepackage{natbib}

\makeatletter
\def\ps@pprintTitle{%
 \let\@oddhead\@empty
 \let\@evenhead\@empty
 \def\@oddfoot{\centerline{\thepage}}%
 \let\@evenfoot\@oddfoot}
\makeatother

\begin{document}

\begin{frontmatter}

\title{On the use of the energy probability distribution zeros in the study of phase transitions}

\author[ufmg]{L A S M\'ol}
\ead{lucasmol@fisica.ufmg.br}
\author[ufmg]{ R G M Rodrigues}
\author[ufmg]{ R A Stancioli}
\author[ufop]{J C S Rocha}
\author[ufmg]{ B V Costa}

\address[ufmg]{ Laborat\'orio de Simula\c c\~ao, Departamento de F\'isica, ICEx, Universidade Federal de Minas Gerais, 31720-901, Belo Horizonte, Minas Gerais, Brazil}
\address[ufop]{Departamento de F\'isica, ICEB, Universidade Federal de Ouro Preto, 35400-000, Ouro Preto, Minas Gerais, Brazil}

\begin{abstract}
This contribution is devoted to cover some technical aspects related to the use of the recently proposed energy probability distribution zeros in the study of phase transitions. This method is based on the partial knowledge of the partition function zeros and has been shown to be extremely efficient to precisely locate phase transition temperatures. It is based on an iterative method in such a way that the transition temperature can be approached at will. The iterative method will be detailed and some convergence issues that has been observed in its application to the 2D Ising model and to an artificial spin ice model will be shown, together with  ways to circumvent them.
\end{abstract}

\end{frontmatter}

\section{Introduction}

Phase transitions are a core subject in statistical physics. Although the concept of phase itself is somewhat ill-defined, phase transitions are much better stated. Indeed, phases can only be clearly defined when phase boundaries or phase coexistence are present, allowing one to identify the differences between them. Phase transitions, on the other hand, are characterized by a non-analytic behavior of the free energy, which in general leads to divergences and discontinuities in thermodynamic quantities that can be ``easily'' identified. However, given the interactions among the elementary degrees of freedom of a system, the identification and characterization of possible phase transitions is not an easy task. As is well known, only very few systems can be exactly solved, in such a way that approximated methods are necessary. Among many possibilities, computer simulations based on the Monte Carlo method\cite{livro-landau} together with finite size scaling theory\cite{livro-privman} can give very accurate results, being thus the preferred method for many studies. 

In computer simulations only finite systems can be studied and since phase transitions only exist in the thermodynamic limit, the use of finite size scaling theory to characterize phase transitions is mandatory.  In fact, as the free energy is obtained by the natural logarithm of the partition function and the latter is a sum of positive terms, non-analyticities can only appear in the free energy density when the thermodynamic limit is taken. Nevertheless, pseudo-phase transitions are present in finite systems and play a very important role in modern physics. A key example are conformational transitions that occur in macromolecules \cite{livro-bachmann}. Polymers are intrinsically finite systems, specially biological polymers (e.g. proteins) in which a different number or arrangement of monomers imply a completely different biological function. Protein folding, for example, can be thought of as a pseudo-phase transition between unfolded and folded states and its comprehension is of extreme importance. Then, methods that are devoted not only to study phase transitions in thermodynamic limit but also to properly explore pseudo-transitions in finite systems are of key importance. 

Recently\cite{Costa17} some of us proposed to use the partial knowledge of the energy probability distribution function to obtain a set of zeros from which phase transitions and pseudo-phase transitions can be identified. Among its advantages, the fact that the transition temperature can be approached at will and that ambiguities related to the location of a pseudo-transition temperature in finite systems are not present are of paramount importance. Actually, when considering pseudo-transitions in finite systems each  thermodynamic quantity signals the transition in a slightly different temperature, making its proper location ambiguous. By using the zeros of the energy probability distribution, such ambiguity is removed, leading to a single estimate of the pseudo-transition temperature for a given system. In what follows we briefly describe the method and detail its use and application. Some convergence issues are also shown together with  ways to circumvent it. We then finish by presenting our closing remarks

\section{Energy probability distribution zeros}

As already mentioned, phase transitions are characterized by a non-analytic behavior of the free energy. This behavior can be captured by considering the complex zeros of the partition function. Indeed, Yang and Lee \cite{YangLee} have shown that the partition function zeros in the complex fugacity plane contain all the relevant information about the thermodynamic behavior of  a system. In particular they showed that in the thermodynamic limit the density of zeros completely determine its critical behavior. In 1964 Fisher \cite{Fisher} extended their idea to the complex temperature plane (Fisher zeros). In summary, by considering the analytic continuation of the temperature to the complex plane, the partition function can be written as a polynomial in a variable $z\equiv e^{-\beta \epsilon}$, where $\beta = 1/k_BT$ is the inverse temperature and $\epsilon$ is a proper energy interval, whose roots contain all thermodynamic information. Formally, the partition function can be written as
\begin{equation}
\label{Z1}
Z=\sum_E g(E) e^{-\beta E}=e^{-\beta \epsilon_0}\sum_n g_n \left ( e^{-\beta \epsilon}\right )^n=e^{-\beta \epsilon_0}\sum_n g_n z^n,
\end{equation}
where $g(E)$ is the number of states with energy $E$ and $E$ can be set to $E=\epsilon_0 + n\epsilon$. Once the above polynomial can be factorized in a set of complex roots, $z_n$, all information present in the partition function is encoded in this set. In particular, the expected non-analytic behavior of the free energy, $F=-k_BT\ln Z$, in the vicinity of a phase transition is expected to be recovered by the presence of a real positive zero, i.e., $Z=0$ for $T=T_c$ in a phase transition. This zero has small imaginary part in finite systems and touches the real axis in the thermodynamic limit. It is called dominant or leading zero. 

The main problems in the Fisher zeros approach are the requirement of the entire density of states, $g(E)$, the high polynomial degree (the number of energy levels), and the elevated values of the coefficients of the polynomial, making numerical methods to find the roots unreliable.
Actually, the polynomial degree easily surpasses 50,000 while g(E) may span over more than a hundred thousand orders of magnitude. In this scenario, there is no efficient numerical method to determine the partition function zeros. In order to circumvent this difficulty, in Ref.~\cite{Costa17}, some of us showed that by rescaling the zeros, the polynomial degree and the range of coefficients can be safely reduced without relevant modifications in the position of the leading zero. 

The proposed polynomial\cite{Costa17} can be obtained by multiplying eq.~\ref{Z1} by $1=e^{-\beta_0 E}e^{\beta_0 E}$. Then,
\begin{equation}
\label{Z1}
Z_{\beta_0}=\sum_E g(E) e^{-\beta_0 E}e^{-(\beta -\beta_0) E}=e^{-\beta \epsilon_0}\sum_n (H_{\beta_0})_n \left ( e^{-\Delta \beta \epsilon}\right )^n=e^{-\beta \epsilon_0}\sum_n (H_{\beta_0})_n x^n,
\end{equation}
where $x\equiv e^{-(\beta-\beta_0)\epsilon}$ and $H_{\beta_0}(E)=g(E)e^{-\beta_0E}$ is the non-normalized energy probability distribution function at (inverse) temperature $\beta_0$. Since $x=e^{\beta_0 \epsilon}z$, up to this point only a rescaling of the Fisher zeros was done. The key point of the method\cite{Costa17} is that now we have a clear criterion to filter the most relevant zeros, specially those that indicate a phase transition. For temperatures near the transition temperature, i.e. for $\beta \approx \beta_c$, states with very low probability to occur are not expected to play any significant role in the transition. Then, one might expect that they do not contribute appreciably to the location of the leading zero, in such a way that even discarding those energy states precise estimates for the leading zero can be obtained. 

\begin{figure}[ht!]
\includegraphics[width=0.47\linewidth]{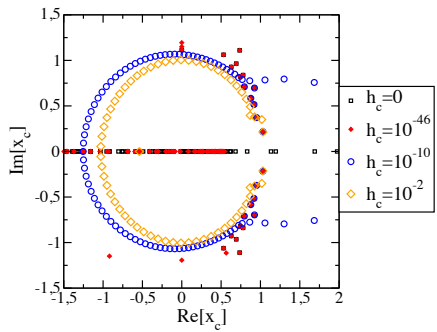}\hspace{0.05\linewidth}%
\begin{minipage}[b]{0.47\linewidth}\caption{\label{corte} Energy probability distribution (EPD) zeros for the $16\times16$  2D Ising model at $\beta_0=0.44$ considering different cut-offs in the histogram, $(H_{\beta_0})_n$ as indicated in the legend. As can be seen, the polynomial order is greatly reduced without noticeable modifications to the location of the leading zero.}
\end{minipage}
\end{figure}

To illustrate that the polynomial can be safely filtered, we refer to Fig. \ref{corte}. This figure was obtained using exact results\cite{Beale} for the density of states for the 2D Ising model in a $16\times16$ square lattice to build the EPD at $\beta_0=0.44$. The black squares are the rescaled Fisher zeros, i.e. the energy probability distribution (EPD) zeros considering all states, obtained using Mathematica\textsuperscript{\textregistered}. The polynomial degree in this case is 257 and the coefficients span over 76 orders of magnitude! As can be seen, the solver fails to find all roots. As all coefficients are real positive numbers, there should be no root in the positive real axis, but many of them were obtained. Then, setting the maximum value of the coefficients (histogram values) to 1 and discarding values smaller than a given threshold, $10^{-2}$, for example, the polynomial degree and coefficients range are drastically reduced. This result is shown by orange diamonds in Fig. \ref{corte}. The polynomial degree was reduced to 63 and coefficients span over only 2 orders of magnitude. Interesting is the fact that the zero that approaches the most the real positive axis is at the same position for all cut-offs shown. This is the leading zero.

In addition to the above discussion, note that for $\beta_0=\beta_c$, where $\beta_c$ is the critical (inverse) temperature, the leading zero ($x\equiv e^{-(\beta-\beta_0)\epsilon}$) would be at the point $(1,0)$ in the thermodynamic limit. For finite systems, a small imaginary part is expected. In addition, for $\beta_0$ not too close to $\beta_c$ further deviations are expected since states relevant to the phase transition are not properly sampled. Then, as the critical temperature is approached, initial guesses for $\beta_c$ can be improved. This allows the application of the following algorithm\cite{Costa17}  to locate the transition temperature:
\begin{enumerate}
    \item{Build a single histogram $ H_{\beta_0^j}$ at $\beta_0^j$ and apply a cut-off if needed.}
    \item{Find the zeros of the polynomial with coefficients given by $ H_{\beta_0^j}$.}
    \item{Find the dominant zero, $ x_c^j$.}
	\subitem a) {If $ x_c^j$ is close enough to the point $(1,0)$, stop.}
	\subitem b) {Else, make $\beta_0^{j+1}=-\frac{\ln \left (\Re e[x_c^j] \right ) }{\epsilon}+\beta_0^j$ and go back to (i).}
\end{enumerate}
In step (iii), the dominant zero can be found, for example, by finding the zero nearest to the point $(1,0)$. As shown in Ref. \cite{Costa17} this algorithm successfully finds the transition temperature for continuous, discontinuous and BKT phase transitions.

\section{Application details}

The use of the above algorithm to study any system where the EPD can be estimated at a given temperature is straightforward. To illustrate it in more detail, consider the 3-state Potts model\cite{Wu} in a square lattice, in which at each lattice site, $i$, a discrete spin variable, $\sigma_i=1,2,3$ is defined. Interactions among neighboring spins are given by:
\begin{equation}
H=-J\sum_{<i,j>}\delta_{\sigma_i,\sigma_j},
\end{equation}
where $J$ is a coupling constant and $\delta_{\sigma_i,\sigma_j}=1$ if $\sigma_i=\sigma_j$ and $\delta_{\sigma_i,\sigma_j}=0$ otherwise. This model has a second order phase transition at $T_c=\frac{1}{\ln(1+\sqrt{3})}\frac{J}{k_B}$ and critical exponent $\nu=6/5$ (see Ref. \cite{Wu}). Good estimates of the EPD can be obtained by conventional Monte Carlo simulations. Here we show results from a simulation using the Metropolis algorithm\cite{livro-landau} for a 20x20 lattice with $5\times10^4$ Monte Carlo Steps (MCS) for thermalization and $5\times10^5$ MCS to build the EPD. The latter is easily obtained in a simulation by counting how many times each energy state is visited and dividing the result by its largest value in order to obtain a normalized histogram. At this point, one could choose to use all visited energy values or cut off the EPD tails since they are, in general, not very well sampled. Remember that since we deal with a finite number of MCS, not all possible energy values are visited in a simulation, in such a way that the energy range (polynomial degree) is naturally reduced as compared to the Fisher zeros. However, in order to prevent possible deviations that may arise due to the poorly sampled tails, we opted to apply a cut-off, discarding values smaller than $10^{-3}$. Care should be taken in this process to ensure that all energy values in the considered energy range, $E\in [E_{min},E_{max}]$, are properly treated. In fact, it may happen to have an intermediate energy value, $E_{min} < E_a < E_{max}$, whose EPD value is smaller than the chosen threshold, i.e., $H(E_a)<10^{-3}$. We remark that $H(E_a)$ must be kept as a polynomial coefficient in order to obtain correct results.
For more details see Ref. \cite{Costa17} Fig. 2. Very similar results are obtained when no cut-off is used and all energy values between the lowest and highest energies sampled in a Monte Carlo simulation are used.

The animated GIF that can be found in the Supplementary Materials illustrates the entire process that leads to an estimate of the critical temperature. We start at $T_0=2 J/k_B$ ($\beta_0=1/2$), building a normalized histogram and cutting off its tails as described above. Using the histogram values as the polynomial coefficients and numerically solving the polynomial, the leading zero is found to be the zero nearest to the point $(1,0)$. From the real part of the leading zero, the new estimate of the inverse critical temperature, $\beta_1$, is obtained from
\begin{equation}
\beta_1=-\frac{\ln \left (\Re e[x_c] \right ) }{\epsilon}+\beta_0,
\end{equation}
where $\Re e\left \{x_c \right \}$ is the real part of the leading zero and $\epsilon$ is the energy step ($E=\epsilon_0+n\epsilon$). Then, a new simulation is done at $T_1=1/\beta_1$, leading to a new histogram from which a new estimate for the critical temperature can be obtained. This process is repeated until the real part of the leading zero is close enough to 1. How close to 1 one should stop depends on the desired accuracy for the critical temperature. The resulting accuracy also depends on the accuracy of the EPD estimate, on the cut-off size and on the zeros finder accuracy. It seems that the EPD quality is the main quantity to focus on in order to improve the critical temperature estimate.

\subsection{Convergence issues}

Solving a high degree polynomial equation is a complicated task even for the most sophisticated algorithms available, e.g., LAPACK and Mathematica\textsuperscript{\textregistered}. As shown in figure \ref{corte}, problems were found even for a relatively small polynomial (degree 257). In addition, even using the EPD technique to filter the most 
relevant zeros, it is still possible to encounter  very high degree polynomials, since the energy range grows with the system volume. Hence, a loss of accuracy in the dominant zero location may be expected for large systems, which may cause a few problems for the algorithm in its simplest form to find the transition temperature. The remaining of this section is devoted to
show very specific cases in which this loss of accuracy was found to cause problems.



\subsubsection{2D Ising model}

\begin{figure} [!ht]
\includegraphics[angle=0, width=0.47\textwidth]{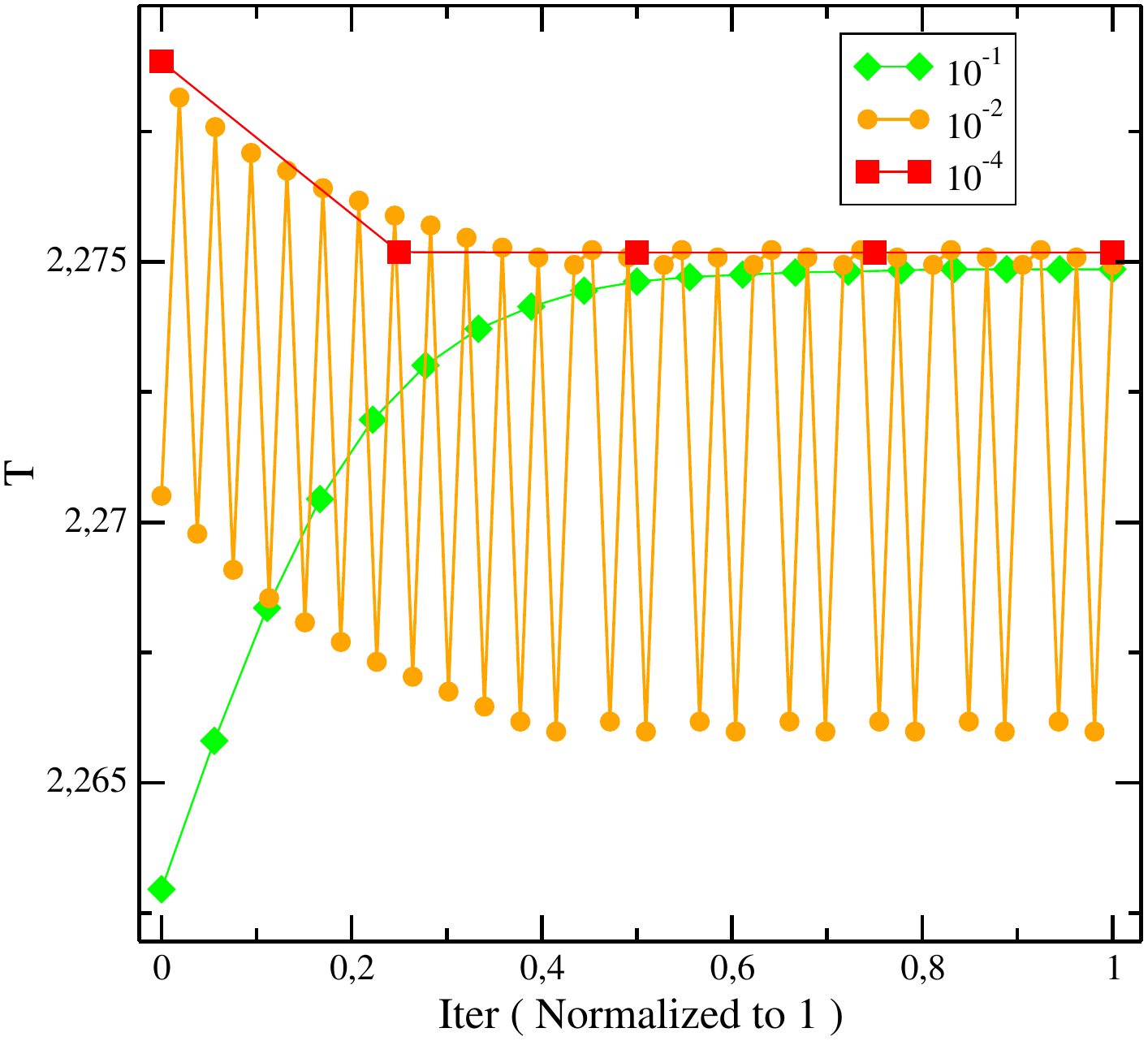}\hspace{0.05\linewidth}
\begin{minipage}[b]{0.47\linewidth}
\caption{\label{conv_t} Convergence towards the critical temperature for the 2D Ising model using the values $10^{-1}$, $10^{-2}$ and $10^{-4}$ as  threshold and conventional Monte Carlo simulations. For the $10^{-2}$ threshold the algorithm did not converge, instead, it starts to oscillate between two branches.}
\end{minipage} 
\end{figure}

Here we report results of conventional Monte Carlo simulations for the 2D Ising model in a square lattice. The simulation was carried out using single spin flips, Metropolis algorithm and $10^8$ MCS at three different
temperatures (2.26, 2.27 and 2.28) in order to use multiple histogram reweighting\cite{Ferrenberg}. For a $150\times 150$ lattice and a $10^{-2}$ threshold, the algorithm previously proposed does not fully converge, instead, the estimated critical temperature starts to oscillate between two different branches as shown in Fig. \ref{conv_t}. Notice that, for the exact
same set of histograms, the algorithm converges, as expected, if we choose the threshold $10^{-1}$ or $10^{-4}$, although a small difference between the two critical temperature estimates can be noticed. The observed difference is expected due to the loss of accuracy related to the different cut-off values used.

\begin{figure} [!ht]
\includegraphics[angle=0, width=0.47\textwidth]{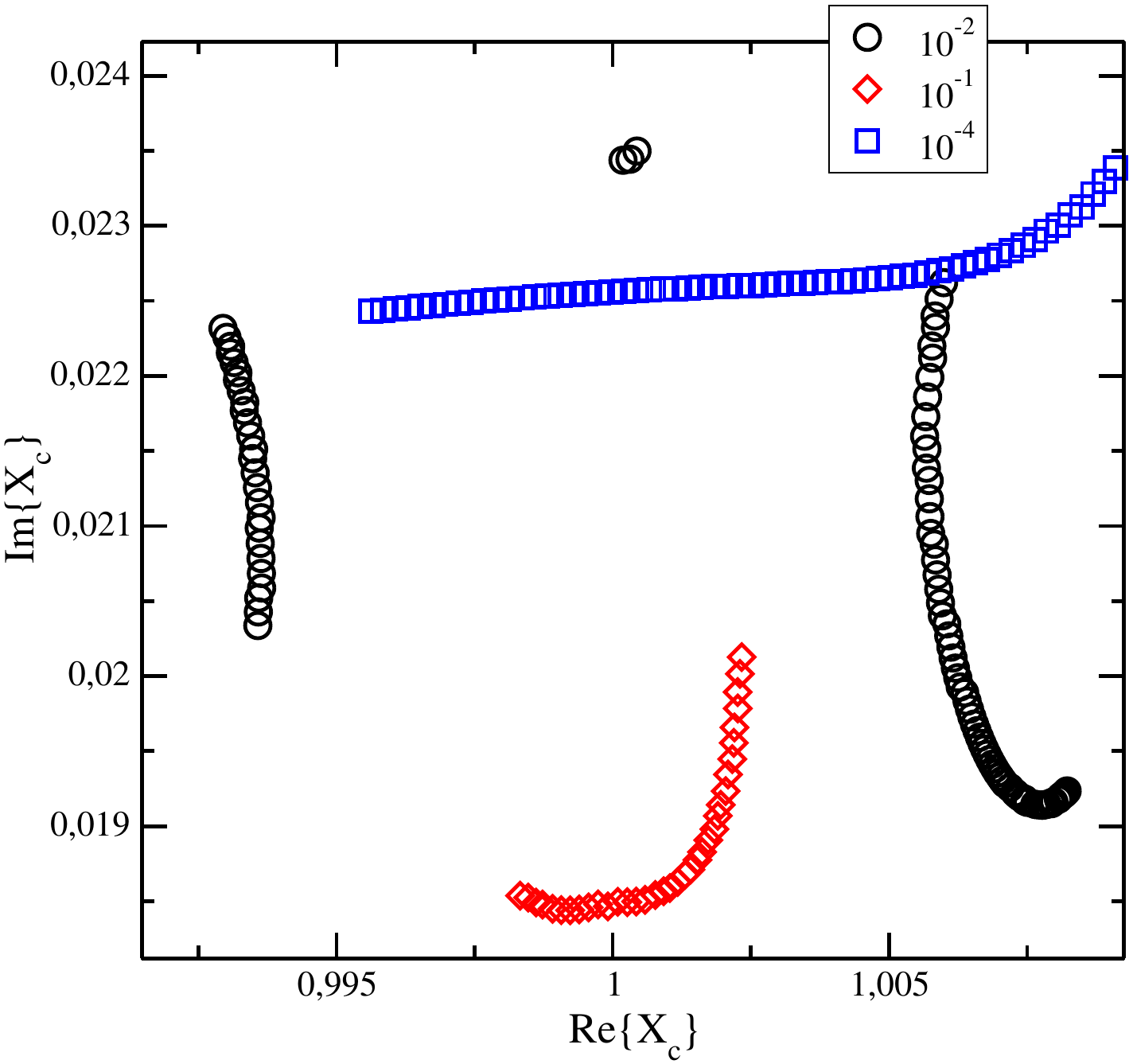}\hspace{0.05\linewidth}
\begin{minipage}[b]{0.47\linewidth}\caption{\label{all_t} Imaginary and real parts of the EPD leading zero for the 2D Ising model for  temperatures $2.26 < T < 2.28 $, with $\Delta T=10^{-4}$ and three different cut-off values.}
\end{minipage} 
\end{figure}

In Fig. \ref{all_t}, instead of considering the algorithm convergence, we show the real and imaginary parts of the leading zero, i.e., the zero nearest to the point $(1,0)$, for three different thresholds, $10^{-1}$, $10^{-2}$ and $10^{-4}$, and temperatures between 2.26 and 2.28 in steps of $10^{-4}$. As can be seen, for the two cases where proper convergence was observed, $10^{-1}$ and $10^{-4}$, the leading zero follows a continuous path, specially near $\Re e\{x\}=1$. On the other hand, for a threshold of $10^{-2}$, it seems that there are forbidden regions, which were observed to cause the convergence problems.

\begin{figure} [!ht]
\begin{center} 
\includegraphics[angle=0, width=0.46\textwidth]{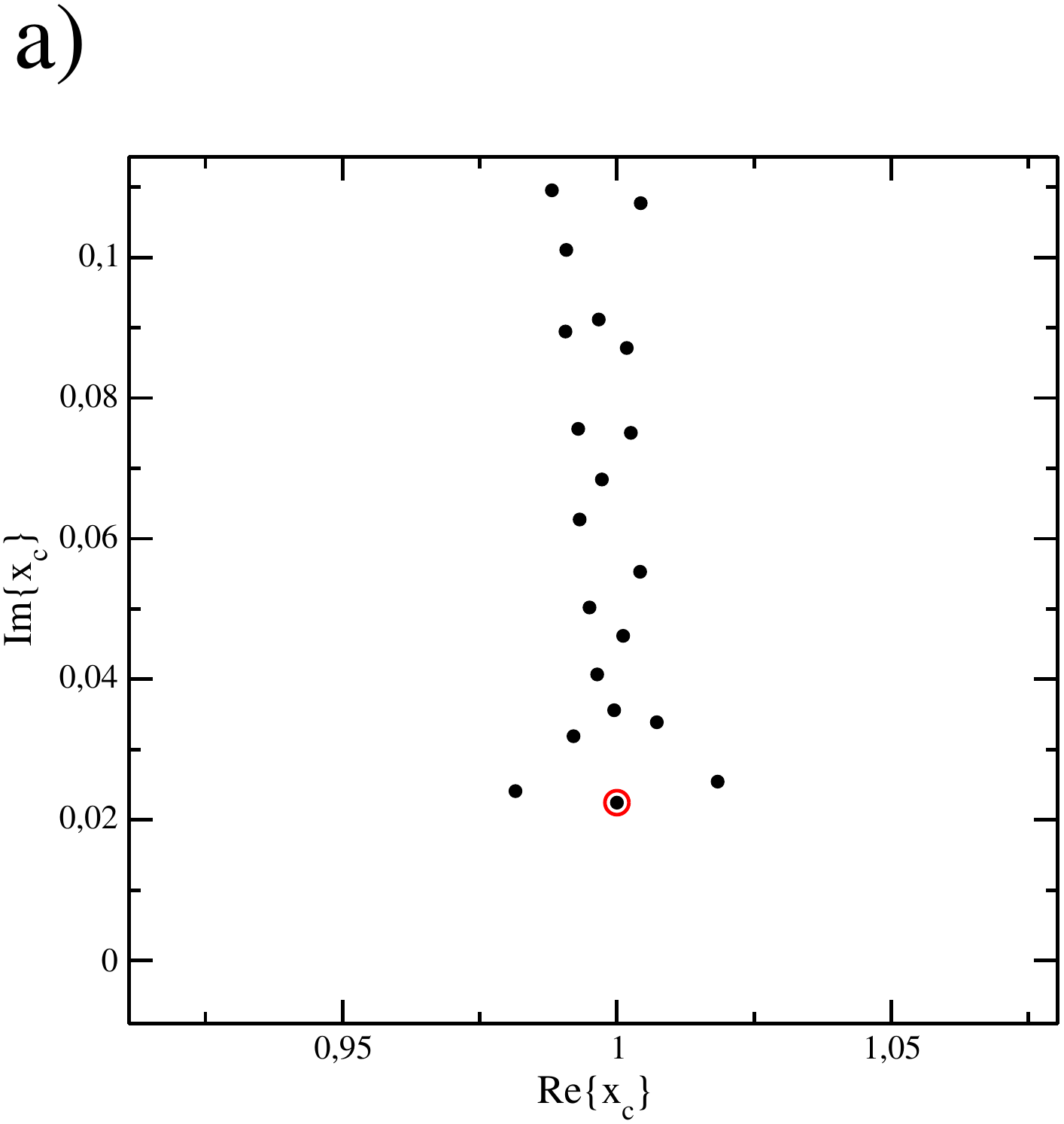}\hspace{0.05\linewidth}
\includegraphics[angle=0, width=0.46\textwidth]{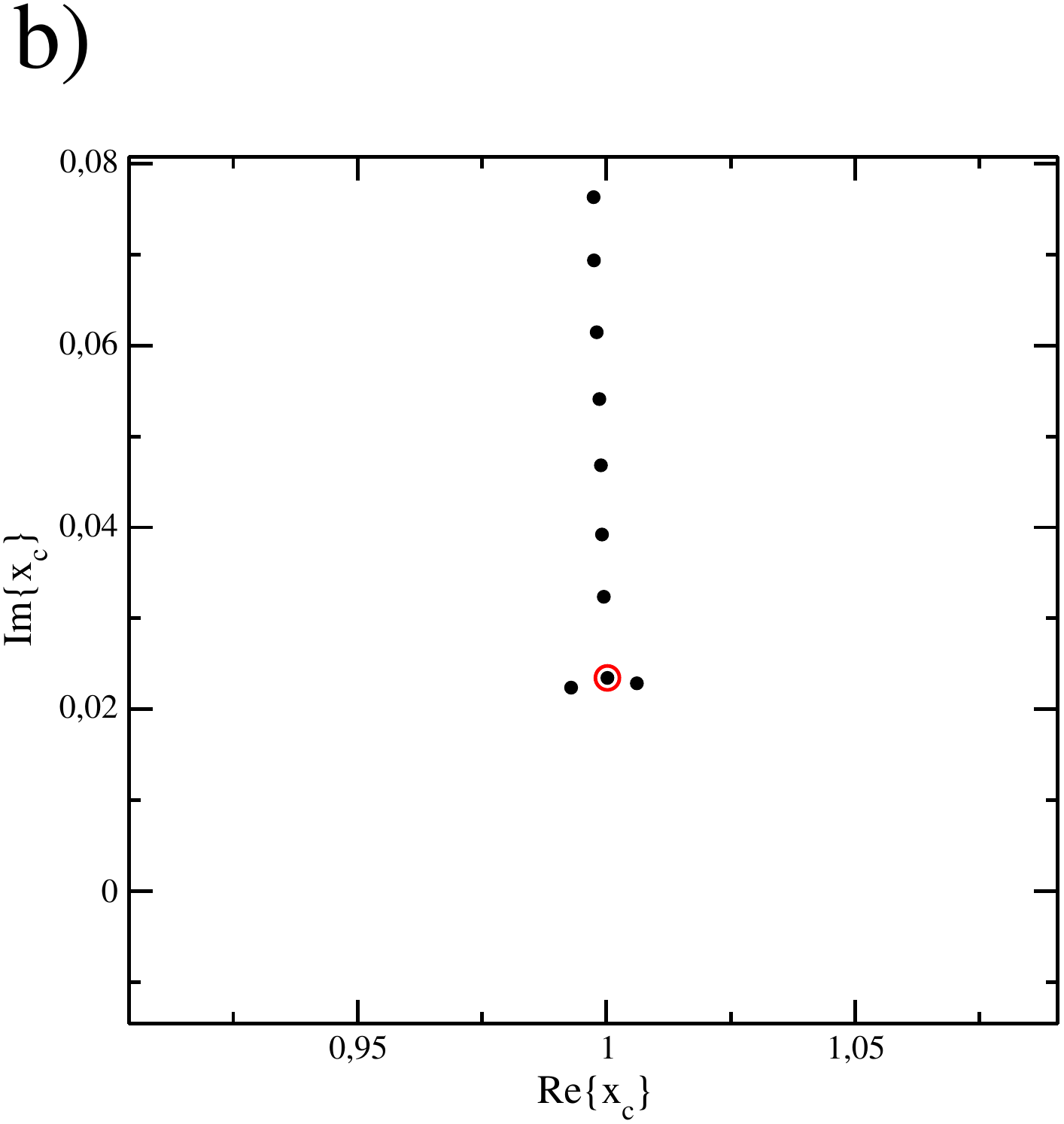}
\includegraphics[angle=0, width=0.46\textwidth]{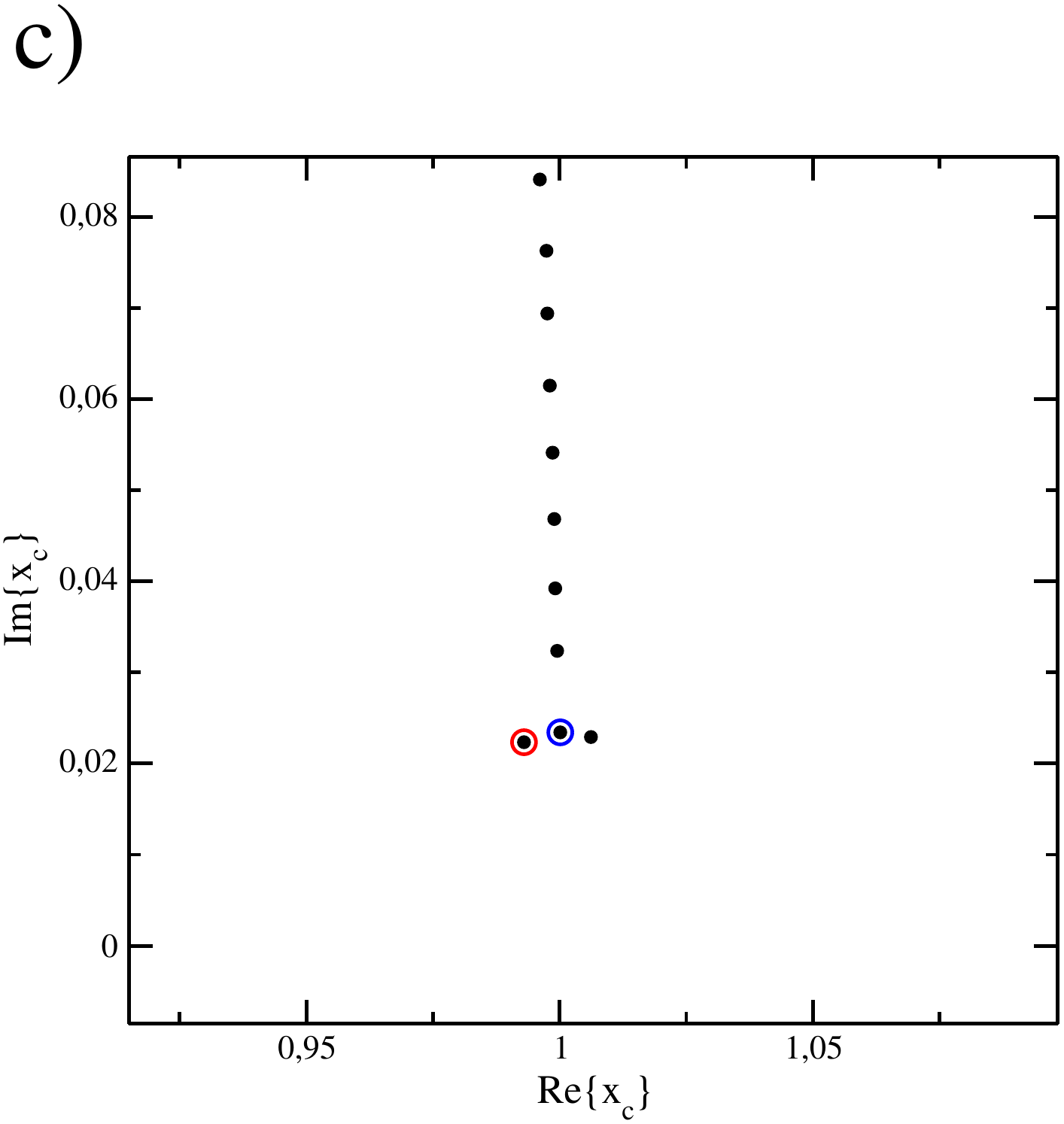} \hspace{0.2cm}
\includegraphics[angle=0, width=0.46\textwidth]{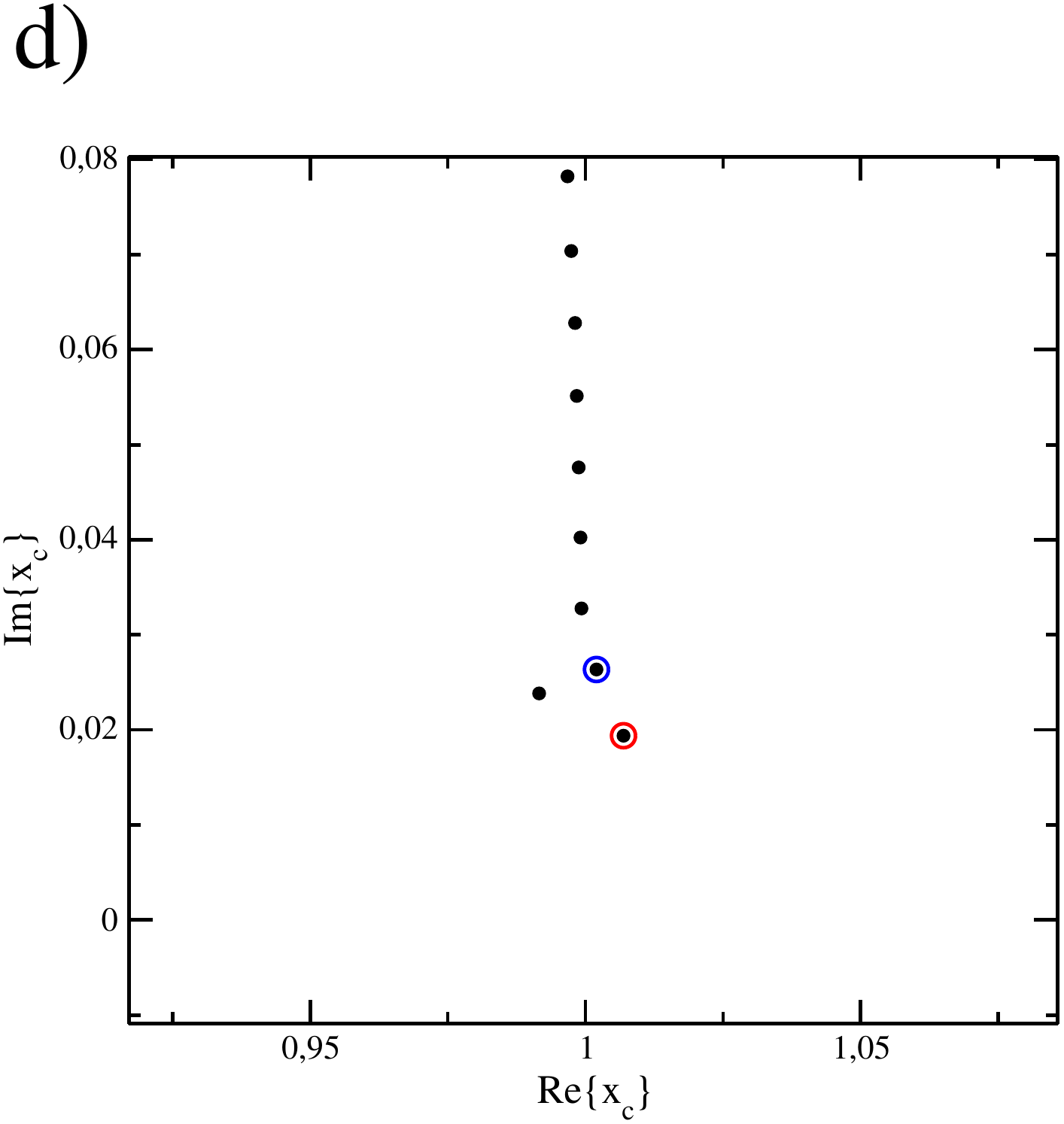}
\caption{\label{roots}  EPD zeros for the $150\times 150$ 2D Ising model. In (a) we show a typical map for a threshold of $10^ {-4}$. In (b), (c) and (d) we show maps for a threshold of $10^{-2}$ at temperatures 
$T=2.2752$,  $T=2.2751$ and $T=2.2662$, respectively. In (b), (c) and (d) we can observe a difference in structure as compared to (a) related to the zeros to the left and to the right of the zero with real part close to 1. }
\end{center} 
\end{figure}

In order to try to understand what is causing this problem, we take a closer look at the zeros map near the point $(1,0)$. We expect a map of zeros at the transition temperature to be like in Fig. \ref{roots} (a), where the leading zero is clearly the closest
to the real axis and to the point $(1,0)$. However, for a $10^{-2}$ threshold, we have a different structure of zeros as shown on the remaining panels of the same figure. Actually, Fig. \ref{roots} (b) is representative of the small cluster of zeros shown near $\Re e\{x\}=1$ for the $10^{-2}$ threshold in Fig. \ref{all_t}, while in (c) and (d) we show typical maps for the left and right branches of  Fig.\ref{all_t}, respectively. Close inspection of these figures suggest that the leading zero may not be the zero nearest to the point $(1,0)$ (marked by red circles). Perhaps, the ``true'' leading zero may be that marked by blue circles in Fig. \ref{roots} (c) and (d). We believe that what is causing this misleading choice of zero are
little fluctuations on the histogram, specially at its tails. A small variation in a coefficient of a polynomial can lead to huge changes on its zeros location. Other possibility may be a poor accuracy of the zeros finder, which may lead to similar results.
Taking into consideration that we have three zeros very close to the point $(1,0)$, even a small perturbation can affect the choice of the leading zero when using the sole criterion of distance to the point (1,0), 
preventing proper convergence of the algorithm. We remark, however, that instead of automatically choosing as the leading zero the zero nearest to the point $(1,0)$, marked with red circles in Fig. \ref{roots}, one can closely inspect the map of zeros and infer that the leading zero would be the one marked by blue circles. By doing that, proper convergence was recovered in this case.

\subsubsection{Artificial spin ice}

A similar behavior has been observed when applying the method to a completely different system, namely an artificial spin ice model. Artificial spin ices \cite{Nisoli13} are arrays of single-domain, enlongated magnetic nanoislands designed to mimic the geometric frustration found in pyrochlore spin ice materials \cite{Harris1997}, which present interesting features such as magnetic monopole-like excitations \cite{Castelnovo2008,Mol09,Mol10}. The interaction between islands is essentially dipolar, and each of them behaves as an effective Ising-like spin, since its magnetic moment is constrained by shape anisotropy to point along the long axis.

\begin{figure}[ht!]
\includegraphics[width=0.35\linewidth]{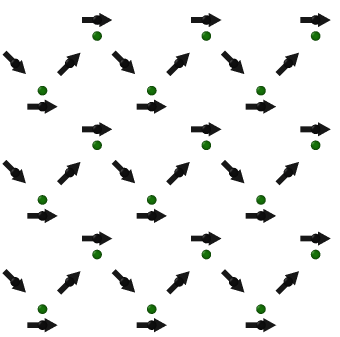}\hspace{0.03\linewidth}%
\begin{minipage}[b]{0.6\linewidth}\caption{\label{rede} Illustration of the lattice geometry simulated. The magnetic islands are treated as Ising spins with dipolar interactions. Periodic boundary conditions are used. In this image we show a $6$x$6$ lattice.}
\end{minipage}
\end{figure}

We have performed Monte Carlo simulations of a spin ice model with a particular geometry recently proposed and experimentally realized by Wang \textit{et al.} \cite{Wang2016}, shown in Fig. \ref{rede}. In our algorithm, the magnetic islands are treated as point-like dipoles and each spin interacts with all other spins in the lattice. The hamiltonian is given by 
\begin{equation}
H  =  \sum_{i>j} \left[ \frac{\vec{m}_i  \cdot \vec{m}_j - 3(\vec{m}_i \cdot \hat{r}_{ij})(\vec{m}_j \cdot \hat{r}_{ij})}{{r_{ij}}^3} \right],
\end{equation}
where $\vec{m}_i$ is the magnetic moment of the $i^{th}$ spin (with $\vert \vec{m}_i \vert=\pm1$), $\hat{r}_{ij}$ is the unit vector that points from spin $i$ to spin $j$ and $r_{ij}$ is the distance between them. The results presented here are for periodic boundary conditions and a lattice size of $32$x$32$ spins.

\begin{figure}[h]
\begin{minipage}{0.47\linewidth}
\includegraphics[width=\linewidth]{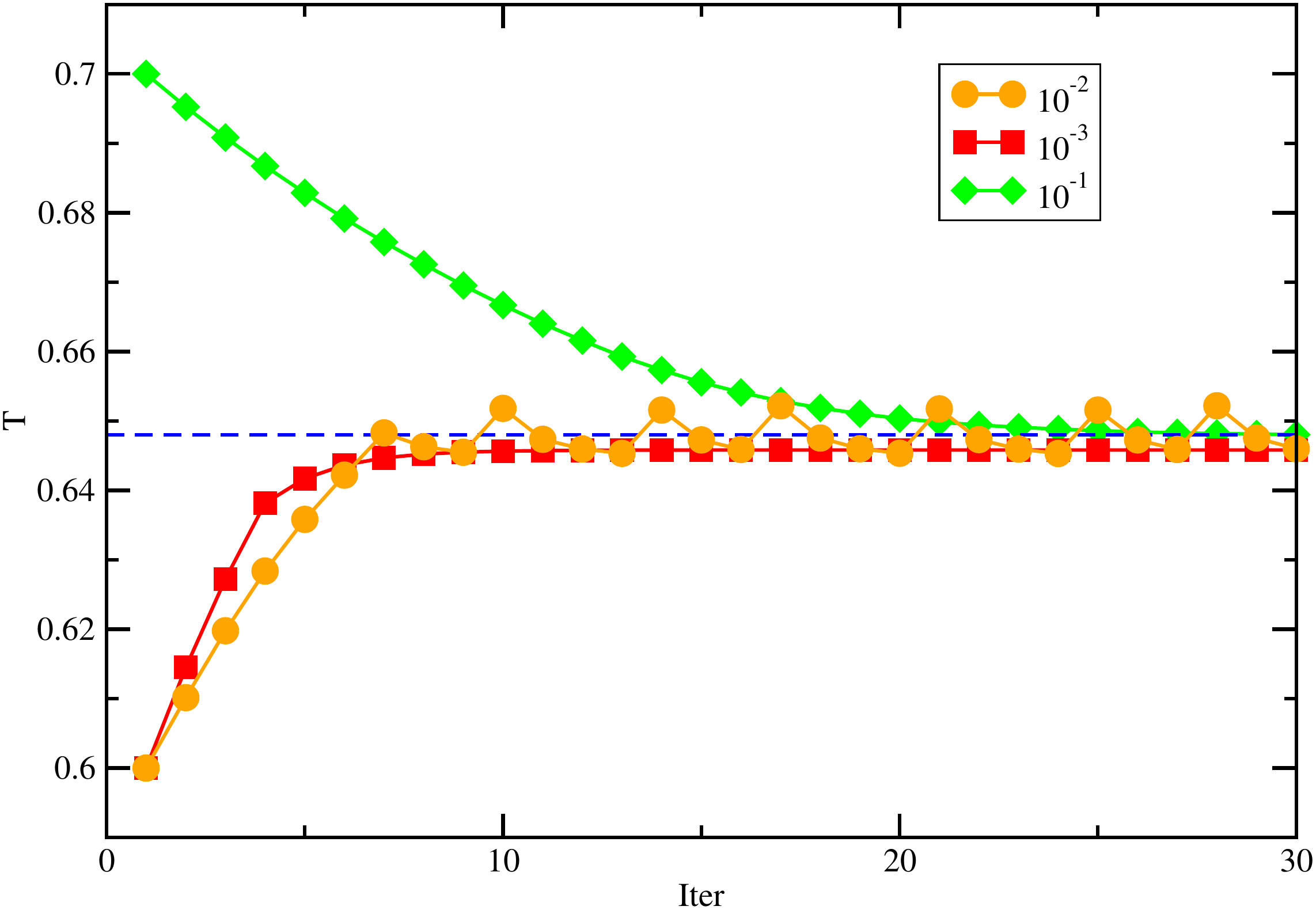}
\caption{\label{temp} Temperature indicated by the leading zero at each iteration of the algorithm for different histogram cut-off values. For comparison, the horizontal dashed line marks the critical temperature estimated by observing the specific heat peak. The choice of starting temperature does not significantly change the results.}
\end{minipage}\hspace{0.03\linewidth}%
\begin{minipage}{0.47\linewidth}
\includegraphics[width=\linewidth]{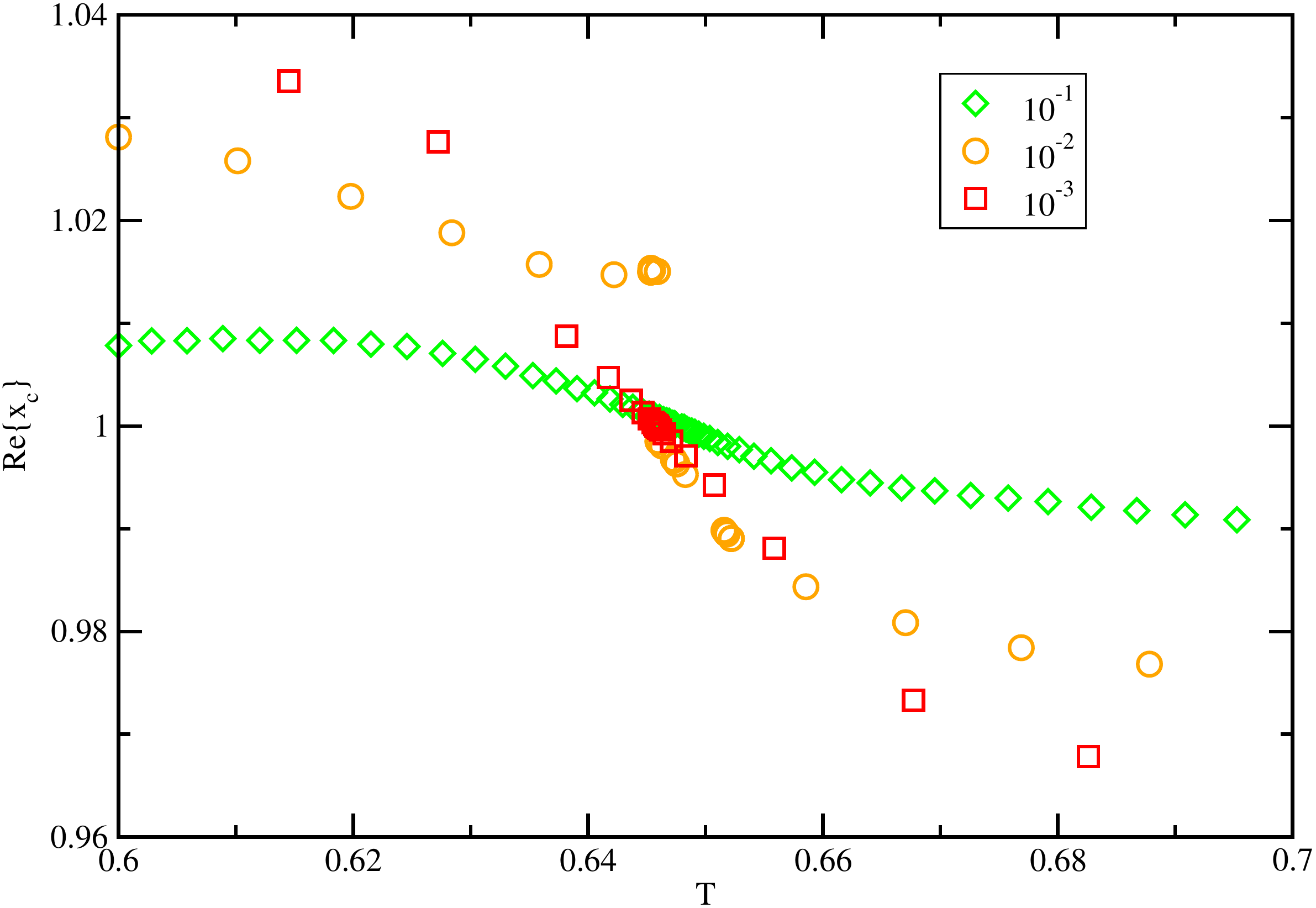}
\caption{\label{realtemp}The real part of the leading zero tends to one as the critical temperature is approached, except for the $10^{-2}$ cut-off, that causes a certain interval on the real axis to be avoided. For each cut-off value, the points shown in the graph are the result of two runs of the algorithm, one starting at a low temperature and the other at a high temperature. We notice that, when the processes converge, the points generated in both runs accumulate near the critical temperature with a real part equal to one. This does not happen with the non-converging processes, in which the leading zeros tend to accumulate elsewhere.}
\end{minipage} 
\end{figure}

Once the histograms had been obtained, three different cut-off values were used to construct the polynomials and solve for the leading zero. As with the Ising model, one of such cut-off values ($10^{-2}$) resulted in the algorithm not converging. Instead, as the expected critical temperature is approached, the leading zero drives the algorithm away from it, producing the oscillating pattern observed in Fig. \ref{temp}. On the other hand, when the cut-off value is set to $10^{-1}$ or $10^{-3}$ the algorithm converges, even though the critical temperature indicated is not exactly the same in both cases, as can be expected due to the different accuracy. Fig. \ref{realtemp} illustrates the convergence process. 
For most cut-off values the real part of the leading zero can get arbitrarily close to one as we approach criticality. Nevertheless, with the $10^{-2}$ cut-off the leading zeros seem to avoid a certain interval on the real axis, which causes the oscillation shown in Fig. \ref{temp}.

\begin{figure}[h]
\includegraphics[width=0.47\linewidth]{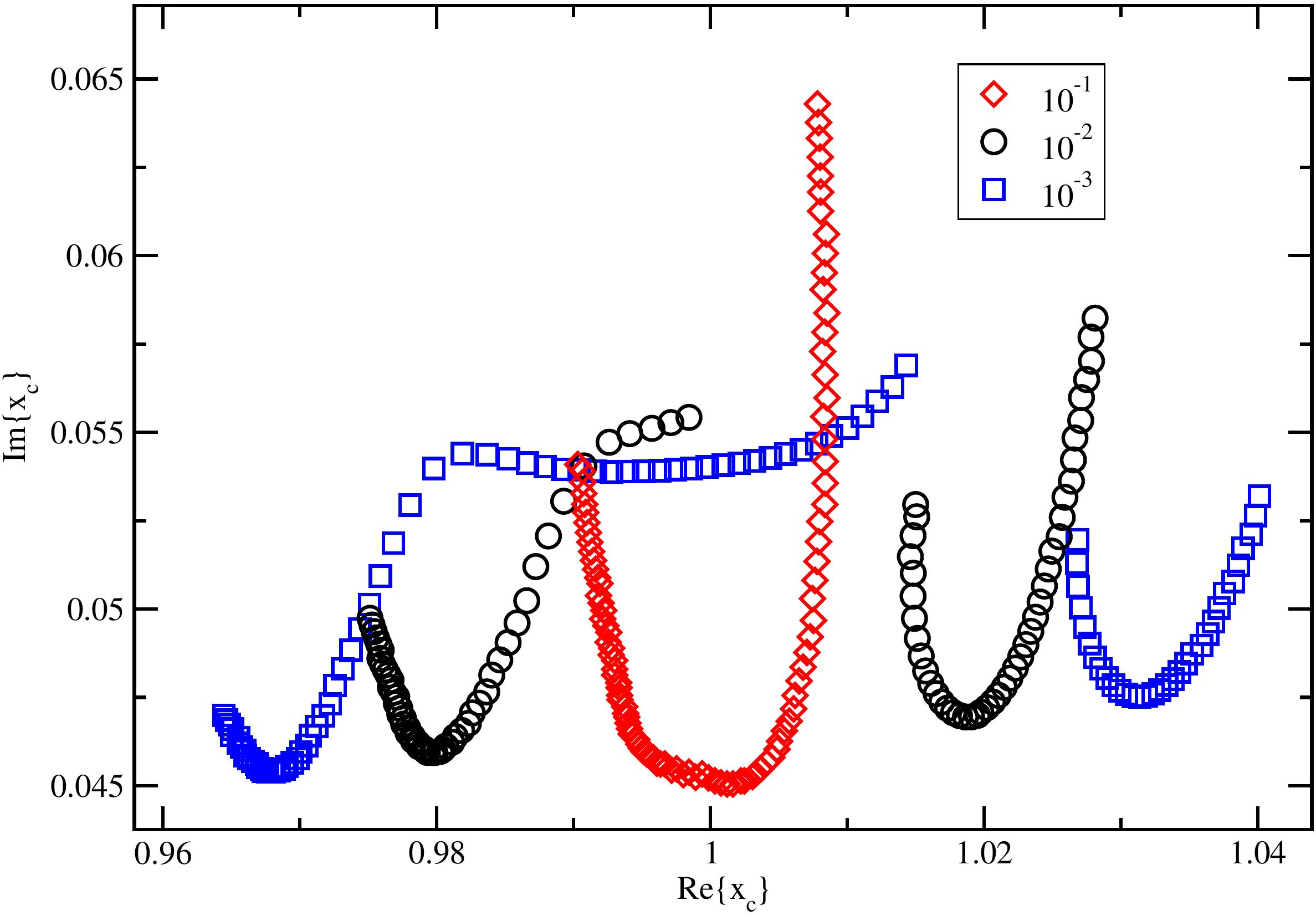}\hspace{0.03\linewidth}%
\begin{minipage}[b]{0.5\linewidth}\caption{\label{varre} Distribution of the leading zeros in the complex plane for each cut-off value. We have scanned temperatures in the interval $0.6<T<0.7$, in steps of $10^{-3}$.}
\end{minipage}
\end{figure}

Once again, we systematically searched for the leading zeros at a certain range of temperatures around criticality. The distribution of the zeros on the complex plane is shown in Fig. \ref{varre} and is very similar to what had been observed for the Ising model (Fig. \ref{all_t}). For the $10^{-2}$ cut-off, a gap appears near $Re[X_c]=1$, while other cut-off values seem to produce zeros that describe a continuous path in this region. In addition, we have found that the maps of zeros are quite similar to that shown in Fig. \ref{roots}. Once more, we remark that proper convergence of the algorithm was observed when instead of choosing as the leading zero the zero nearest to the point $(1,0)$ a careful analysis of the zeros map is done.

\section{Closing remarks}

In summary, we have shown some details concerning the use of the recently proposed zeros of the energy probability distribution\cite{Costa17} in the study of phase transitions. As shown, by using this method there is no need to define an order parameter to find the transition temperature. In addition, there is no ambiguity in the determination of the pseudo-transition temperature for finite systems and the computational effort to locate the transition temperature is small compared to the use of Fisher zeros and other conventional methods. Indeed, since only partial knowledge of the density of states is required and numerical problems related to the zeros finder's task are greatly reduced without noticeable modifications on the leading zero location, we believe that the energy probability distribution zeros constitute a major improvement in the numerical study of phase transitions. Although some convergence problems were found for the 2D Ising model and an artificial spin ice model, they can be solved by proper inspection of the zeros maps and wise definition of the leading zero. Perhaps, a better definition of which zero is the leading zero or improved accuracy of the zeros finder may solve the observed convergence issues. These questions will be properly addressed in the near future.

\section*{Acknowledgments}
The authors thank CNPq, CAPES and FAPEMIG, Brazilian agencies, for partial support.


\bibliography{refs}
\end{document}